\documentclass[pre,twocolumn,a4paper,superscriptaddress]{revtex4}
\usepackage{epsfig}
\usepackage{graphics}
\begin{document}

\title{Template coexistence in prebiotic vesicle models}
\author{Daniel G. M. Silvestre and 
	Jos\'e F. Fontanari \thanks{e-mail: {\tt fontanari@ifsc.usp.br} }
       }                    
\affiliation{Instituto de F\'{\i}sica de S\~ao Carlos,
Universidade de S\~ao Paulo, Caixa Postal 369, 13560-970 S\~ao Carlos, 
S\~ao Paulo, Brazil}

\begin{abstract}
The coexistence of distinct templates is a common feature of the diverse proposals advanced to resolve the information crisis of prebiotic evolution. However, achieving robust template coexistence turned out to be such a difficult demand that only a class of models, the so-called package models, seems to have met it so far. Here we apply Wright's Island formulation of group selection to study the conditions for the coexistence of two distinct template types confined in packages (vesicles) of finite capacity. In particular, we show how selection acting at the level of the vesicles can neutralize the pressures towards the fixation of any one of the template types (random drift) and of the type with higher replication rate (deterministic competition). We give emphasis to the role of the distinct generation times of templates and vesicles as yet another obstacle to coexistence. \end{abstract}
\maketitle
\section{Introduction} \label{sec:Intro}
Though a ubiquitous feature of life (e.g., genomes, multi-cellular organisms and animal societies) cooperation poses a difficult problem to the classical interpretation of evolution by natural selection as an egoistic process in which individual organisms compete fiercely to guarantee the presence of their genes in future generations. Cooperative traits are costly because they presuppose the investment of resources towards a public good, thus benefiting other individuals who may utterly fail to contribute to the community welfare. Traits that benefit the group as a whole but are deleterious to their bearers used to be termed \textit{altruistic} but, probably to avoid the heavy anthropocentric connotation of this term, the modern literature favors the denomination \textit{cooperative} traits instead \cite{Sachs04}. There is an extensive literature on the evolution of the cooperation in nature (see e.g. \cite{Sachs04,Wade78,Boorman80,Wilson80}) but only very recently a series of experiments on microbial populations have substantiated the theoretical arguments supporting it \cite{Rainey03,Velicer03,Sachs05}.

A key element to explain the evolution of cooperative traits is the existence of some form of non-random association between individual members of the population. For instance, cooperation may be restricted to a group of individuals sharing a common ancestor. This is the so-called kin selection mechanism that gained fame by explaining the cooperative behavior of social insects \cite{Hamilton64}. Curiously, although kin selection is fittingly considered a kind of group selection, it relies entirely on a gene's-eye view in which the unit of selection is the gene rather than the individual or the group. A robust alternative to the postulation of mechanisms of recognition between kin individuals is the spatial enclosure or compartmentalization of small groups of individuals since then the offspring
of cooperators will remain close to their relatives and the benefits from cooperation will be confined mainly to the group of cooperators. In fact, compartmentalization is now acknowledged as an obligatory stage in prebiotic evolution needed to enforce cooperation among distinct templates and so to set the conditions for the formation of a gene network necessary to cellular life \cite{Maynard}. Henceforth, we will use interchangeably the terms template and replicator to refer to a self-replicating molecule formed by a sequence of nucleotides.

The awareness of the necessity of a primitive mechanism to impose cooperation among templates has grown from the seminal work of Eigen on purely competitive templates  which came to be known as the quasispecies model \cite{Eigen}. Particularly relevant to our purposes was the finding that the length of a replicating polymer (i.e., a RNA-like template) is limited by the replication accuracy per nucleotide and so primordial replicators would have to replicate with very high accuracy in order to reach the length of today's RNA viruses (about $10^{3}$ to $10^{4}$ nucleotides). Such a replication accuracy, however, cannot be achieved without the aid of specialized catalysts (peptide enzymes), but
building those catalysts requires a blueprint that amounts to a large nucleotide sequence, leading thus to a molecular version of the old puzzle about the chicken or the egg --
no large genome without enzymes, and no enzymes without a large genome \cite{Maynard83}.
This finding and the observation that templates with distinct replication rates cannot coexist \cite{Swetina82} triggered the so-called information crisis of prebiotic evolution. This crisis brought to light the challenging conclusion that, in spite of being at the heart of natural selection, competition alone would not have worked in prebiotic times: some form of cooperation between templates is mandatory to overcome the information crisis.

Cooperation can bypass the information crisis by allowing the coexistence of several short templates, i.e., by splitting the information in  modules, similarly to the division of the genome in chromosomes found in many organisms. To ensure the coexistence of distinct replicators,  Eigen and Schuster proposed a cyclic reaction scheme, termed hypercycle, in which each replicator would aid in the replication of the next one, in a regulatory cycle closing on itself \cite{hyper}. Clearly, this proposal requires that the primordial replicators functioned both as template and replicase. Although the discovery of the  
catalytic activity of RNA has lent credibility to the hypercycle scenario \cite{Doudna89},
the assumption that each replicator has two separate functions, namely, a replicase for the next member of the hypercycle and a target for the replicase associated to the previous member faced strong criticism \cite{Bresch80}. In fact, while it is obvious that natural selection will act so as to make each element of the hypercycle a better target for replication, it will oppose or at least not favor the cooperative part of the scheme, i.e., to make the replicator a better replicase for other replicators. This function is then certain to  degenerate quickly since deletions and mutations that impair it would carry a  selective advantage. In that sense, as Maynard Smith pointed out, giving catalytic support in such molecular networks is an altruistic behavior and so hypercycles are easy targets to parasites, i.e., molecules that do not reciprocate the catalytic support they receive \cite{Maynard79}. The ruin of the hypercycle is then an unavoidable consequence of natural selection.

An alternative suggestion to resolve the problem of the coexistence between templates which is very much in line with the classical works on the origin of life \cite{Oparin52,Fox65} is to enclose the unlinked templates in isolated compartments or vesicles. The key to coexistence is to assume that the vesicles proliferate with a production rate that depends on their template compositions. Essentially, this amounts to assume that the coupling among different template types occurs through a common metabolism which is ultimately the responsible for the survival and reproduction of the vesicle, and that the well-functioning of this metabolism requires the contribution of all template types. This is the central idea behind the so-called package models \cite{Niesert81}, among which the stochastic corrector model \cite{Szathmary87,Grey95,Czaran00,Zintzaras02} is the most popular. As revealed by the word `stochastic', a crucial ingredient of these models is the finitude of the population of templates within each vesicle. In fact, it is the stochastic nature of the template dynamics that produces diversity in the population of vesicles, creating thus the opportunity for the operation of natural selection.

In contrast with the quasispecies and hypercycle models, very little is known about the dynamics and stationary states of package models, since the great complexity resulting from the coupling of template and vesicle dynamics precludes a full analysis of the space of parameters of the models. Such a systematic analysis is important to determine in what conditions, if any, template coexistence can be achieved. The situation here is similar to
that found in models of group selection (see, e.g., \cite{Wade78,Boorman80,Wilson80} for reviews). In particular, the coupling between the two dynamics can be treated analytically provided there is a countable infinity of vesicles, so the dynamics at the group level is deterministic \cite{Eshel72,Aoki82,Silva99}. Otherwise, this group selection formulation retains the main ingredients of the package models in that the number of templates within each vesicle is finite and the  survival and consequent proliferation of the vesicles depends on their template compositions. In this contribution we broaden a preliminary study on the suitability of the classic group selection framework to study coexistence of templates in package models \cite{Fontanari}. In particular, we relax the unfounded but widely used assumption of group selection models (see, e.g., \cite{Eshel72,Aoki82,Silva99,Fontanari}) that templates and vesicles have the same generation times. In doing so we found that coexistence is impossible if the template generation time is much shorter than that of the vesicles and that, when possible, template coexistence is achieved only within a well-defined range of the vesicle capacity.
\section{Model} \label{sec:Model}
Following the formulation of Wright's Island model \cite{Wright} we consider a global population composed of an infinite  number of spatially isolated local populations - the vesicles - each of which encloses exactly $N$ templates. This framework, which forms the foundation for traditional group-selection theories, has been successfully used to study the conditions for the evolution and maintenance of altruistic traits in nature (see, e.g., \cite{Boorman80}). Here we employ the Island model to study a more difficult problem, namely, the coexistence of two distinct template types $A$ and $B$ within a same vesicle of capacity $N$. Without loss of generality we assume that template $A$ has replication rate $1- \tau$ with $\tau \in [0,1]$ and template $B$ replication rate $1$. Hence  $\tau$ is referred to as the handicap parameter. The vesicles are identified by their template compositions or, more pointedly, by the number $i=0, \ldots,N$ of type $A$ templates they wall in. The state of the global population at a given generation $t$ is completely specified by the frequencies of vesicles of type $i$ (i.e., a vesicle that encloses $i$ templates $A$ and $N-i$ templates $B$), denoted by $Y_i^t$ with $\sum_i Y_i^t = 1$ for all $t$. Given the mechanisms for template competition that takes place inside the vesicles and for the competition between vesicles, our goal is to derive a recursion equation for the frequencies of the $N+1$ different vesicle types. The life cycle of the vesicles (i.e., one generation) consists of three events -- vesicle extinction, vesicle recolonization and template replication -- which take place in this order and are described as follows.
\subsection{Vesicle extinction} \label{sec:extinction}
As pointed out in Sect.~\ref{sec:Intro}, the coexistence of distinct template types solves
the information crisis problem because the information content of a vesicle may be seen as
split into several (two, in our case) parts, which must all be present at any time to guarantee the viability of the system as an integrator of information. Moreover, it would be desirable that the different templates contribute with approximately the same number to the vesicle composition. This constraint, namely, the  presence of both template types in equal
concentrations within each vesicle, can be enforced by choosing an appropriate measure for
the survival probability of type $i$ vesicle, which we denote by $\alpha_i \in [0,1]$. Here we choose the geometric mean \cite{Niesert81,Szathmary87,Grey95,Czaran00,Fontanari} 
\begin{equation} \label{geom}
\alpha_i = 1 - g + \frac{2 g}{N} \left [ i \left ( N - i \right ) \right ]^{1/2}  
\end{equation}
where $g \in [0,1]$ is a parameter measuring the benefit brought to the vesicle by the presence of the two template types within it. Hence regardless of the value of $g >0$ survival is guaranteed ($\alpha_i = 1$) for vesicles with an even template composition ($i = N/2$). Vesicles that lack one of the template types, i.e., vesicles of type $i=0$ and $i=N$ are assigned a baseline survival probability $1-g$, so that the selective pressure against these vesicles increases with increasing $g$. We note that the vesicle survival probability given in equation (\ref{geom}) can be seen as describing a dynamical link among the template types through a common metabolism -- each template contribute to the good of the vesicle by catalyzing its metabolism at various points \cite{Szathmary87}. The absence of any of the catalysts would then greatly impair the vesicle metabolism and so its survival capability.
\subsection{Vesicle recolonization} \label{sec:recolonization}
The net result of the extinction procedure described before is that a fraction $1 - \sum_i \alpha_i Y_i$ of vesicles disappear and must then be recolonized, i.e., replaced by the surviving vesicles. This is done by replicating these vesicles in proportion to their frequencies in the population just after the extinction procedure, yielding thus the following new vesicle frequencies
\begin{equation} \label{recol}
\frac{\alpha_i Y_i^t}{\sum_j \alpha_j Y_j^t}
\end{equation}
for $i=0,\ldots,N$. This equation prompts the interpretation of the parameter $\alpha_i$ as the replication rate of a vesicle of type $i$ and so henceforth we will refer to the joint processes of extinction and recolonization as the process of vesicle replication. We note  that this standard procedure for recolonization (see, e.g., \cite{Eshel72,Aoki82,Silva99})
implies the instantaneous replacement of the extinct vesicles by the surviving ones with the probability given in equation (\ref{recol}). For example, in an extreme situation, in which only one vesicle passes the extinction stage, this sole vesicle will replenish the entire population (infinite or finite) in a single time step. Although rarely made explicit, this assumption seems to underlie all deterministic population models with non-overlapping generations. However, this drawback can be safely ignored if $g$ is not close to $1$ which, fortunately, is the relevant situation in  prebiotic evolution (see discussion in  Sect.~\ref{sec:conclusion}).
\subsection{Template replication} \label{sec:replication}
Since the capacity of the vesicles is fixed and finite, it is necessary to use a stochastic approach to model the dynamics of the templates inside each vesicle. As usual, we assume that the number of copies that a template brings forth is proportional to its relative replication rate. In addition, assuming that there is no overlap between consecutive generations of templates, i.e., between the parent and its clones, we can write the probability that a vesicle of type $j$ changes to a vesicle of type $i$ as $R_{ij} = \left( \mathbf{T}^m \right)_{ij}$ where the elements of the matrix $\mathbf{T}$ are given by
\begin{equation}
T_{ij} = \left ( \begin{array}{c} N \\ i \end{array} \right )
\left ( w_j \right)^{i} \left ( 1 - w_j \right )^{N-i} 
\end{equation}
and  $w_j = j(1 - \tau)/(N-j \tau)$ is the relative replication rate of templates of type $A$ in a vesicle of type $j$. Clearly, $\sum_i R_{ij} = 1 ~\forall j$. The new ingredient here is the integer parameter $m \geq 1$ that yields the number of replication cycles each local template population goes through for each generation of the vesicle population. In other words, if we set the generation time of the vesicles to $1$, then the generation time of the templates will be $1/m$, and so we refer to $m$ as the ratio between vesicle and template generation times. Our formulation for the template dynamics is essentially the celebrated Wright-Fisher model of population genetics \cite{Crow70}, which is very well suited to describe stochastic effects (e.g., random drift) in finite populations. Up to now only the extreme case  $m=1$, in which  templates and vesicles have the same generation time,
was considered in the literature \cite{Eshel72,Aoki82,Silva99,Fontanari},
though it is clear that a more plausible scenario would be to consider $m \gg 1$.
We note that in some alternative prebiotic package models, in which the size of the 
local template population is allowed to increase,  
the vesicle replication stage is triggered when the template population reaches a certain 
size which is fixed \textit{a priori} \cite{Niesert81,Szathmary87,Czaran00,Zintzaras02}.
Thus, fixing this limiting size is similar to fixing our parameter $m$.

\subsection{Dynamics} \label{sec:recursion}
Finally, given the events comprising the life cycle of  templates and vesicles we can 
write a recursion equation for the frequencies of vesicles of type 
$i=0,1,\ldots,N$ in the global population
\begin{equation}\label{recursion}
Y_i^{t+1} = \frac{\sum_{j=0}^{N} R_{ij} \alpha_j Y_j^t}{\sum_{j=0}^{N} \alpha_j Y_j^t} .
\end{equation}
Even in the stationary regime $t \to \infty$, a closed solution for $Y_i^{t}$ is not possible,
except when the template and vesicle dynamics decouple, which happens for
$g=0$ and $m \to \infty$. In both cases random drift ensures  that for \textit{finite}
$N$ either  template $A$ or template $B$ will reach fixation within a given vesicle, i.e.,
$Y^\infty_i = 0$ for $i \neq 0,N$. In fact, the very possibility of fixation of the less
fit template $A$ in a few vesicles together with the  increment of the survival 
probability of those vesicles are  key ingredients of the classic models 
for the evolution and stability of altruistic traits \cite{Eshel72,Aoki82}.

We note that our approach, based on Wright's framework for spatially structured populations,  
contrasts with Wilson's formulation \cite{Wilson80} 
in which the group structure is dissolved each generation 
to form a global mating pool (see \cite{Donato96,Donato97}
for application in ecology  and \cite{Michod83,Alves01}  for application
in prebiotic evolution). 
In particular, coexistence is favored in this  transient group formulation 
since different templates in distinct vesicles are likely to be assigned to the same vesicle 
during the group re-assembling procedure after the mating stage. 

In this contribution we do not take into account the possibilities of 
interchange of templates between vesicles (migration) and  
mutations that change template $A$ into $B$ and vice-versa.   
These processes actually promote coexistence by preventing the fixation of 
the templates and so they are important to test the robustness of models for the evolution 
of altruism,  the  aim of which is the fixation of  the less fit template $A$ rather
than the coexistence between the two template types \cite{Silva99}.
In the following sections we will  characterize  the  stationary solutions of 
the  recursion equations (\ref{recursion}) for a  wide range of the control parameters of
the model.

\section{Independent vesicles} \label{sec:independent}
%

\begin{figure}
\resizebox{0.75\columnwidth}{!}{%
  \includegraphics{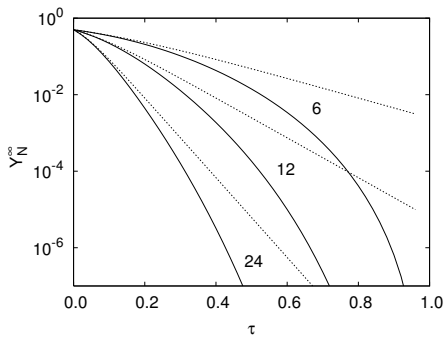}
}
\caption{Semi-logarithm plot of the stationary frequency of vesicles containing
the less fit template  as function of the handicap parameter  
for independent vesicles of capacity $N=6,12$ and $24$. 
The solid lines are the exact results obtained from 
iteration of the recursion equations and the dashed lines are the results of the diffusion
approximation.}
\label{fig:1}       
\end{figure}

In order to better appreciate the many obstacles  hindering the coexistence of distinct
templates confined in  vesicles of finite capacity, in this section we focus on the
simpler problem in which the vesicles evolve independently of each other. As pointed out
before, this is achieved by setting $g=0$ in equation (\ref{geom}) 
so that the survival of the vesicles is guaranteed regardless of their template 
compositions. Alternatively, by letting $m \to \infty$ we allow random drift to fix one
of the templates ($i=0$ or $i=N$) and since $\alpha_0=\alpha_N=1-g$ the competition
between vesicles is effectively turned off. Actually,  the reason there is no competition
in this neutral situation is  because the number of vesicles is infinite. If there were a
finite number of vesicles then 
random drift, now acting at the level of the vesicles,
would  lead again to the fixation of only one vesicle type, in spite of the
fact that both types have the same  survival probability. Use of the diffusion approximation,
valid in the limit of large $N$ and small $\tau$ such that the product $\tau N$ is finite,
yields a simple expression for the fraction of vesicles carrying the less fit
template $A$ \cite{Crow70}
\begin{equation}
Y^{\infty}_{N} = \frac{1 - \exp \left (N \tau p \right )}{1 - \exp \left (N \tau \right )} .
\end{equation}
where $p$ is the initial frequency of template A in each vesicle. Clearly,
$Y_0^\infty = 1 -  Y^\infty_N$.
This analytical prediction is compared with the  results obtained by the numerical iteration
of the recursions (\ref{recursion}) in Figure \ref{fig:1}, where we have used
$Y_{N/2}^0 = 1$ (hence $Y_i^0 = 0$ if $i \neq N/2$) so that $p=1/2$. As expected, there is a good
agreement between the exact numerical and the approximate analytical results for small 
values of the handicap $\tau$. In particular, for $\tau =0$ we have $Y_N^\infty = p = 1/2$.
However,  we find that  $Y_N^\infty$ decreases much faster than 
$\exp (-N\tau/2)$ with  increasing  $\tau$.

The point here is to stress that although random drift is a 
key element of models for the evolution of altruistic traits, as it enables
the fixation of the less fit template in some vesicles (provided the handicap $\tau$ is 
small), it is a serious hindrance to the coexistence of distinct templates within
a same vesicle. It is in this sense that we  can say the  coexistence issue  is more
tricky  to explain (and actually much less studied) than the problem of the altruism in nature.

\section{Deterministic limit} \label{sec:deterministic}

Another hindrance to  template coexistence, as well as to the evolution of altruism via the
fixation of the less fit template $A$, is the deterministic pressure in favor
of the fitter template $B$ that prevails in the limit of infinitely large vesicle capacities
$N \to \infty$, regardless of the value of the cooperation pressure $g <1$. In fact, the 
reason that the uniform initial vesicle population (i.e., all vesicles composed of the same 
number of templates $A$ and $B$)
used to draw Figure $\ref{fig:1}$
resulted in  the two antagonistic types of vesicles was the amplification of random fluctuations
which  is an inherent feature of the competitive dynamics in a finite population. 
For $N \to \infty$  such fluctuations are
absent and so all  vesicles have the same composition at all generations. As a result, 
the competition between vesicles is turned off thus leading to the fixation of the fitter 
template in all vesicles (except in the degenerate case $\tau=0$,  for which the two
template types coexist forever).

It is clear then that template coexistence will be possible only within a narrow range
of the values of the control parameters and, in particular, of the vesicle capacity. 
In the following section we determine the regions in the parameters space where
coexistence is possible.

\section{Template coexistence} \label{sec:coexistence}
%

\begin{figure}
\resizebox{0.75\columnwidth}{!}{%
  \includegraphics{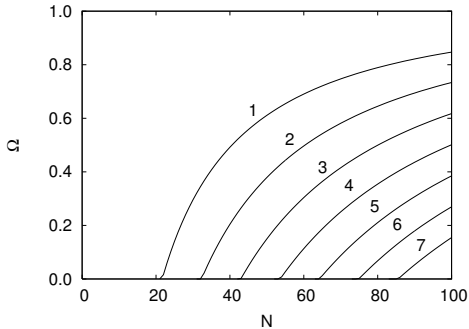}
}
\caption{Fraction of vesicles with the two template types in their compositions 
as function of the vesicle capacity for $\tau=0$ (i.e., the templates have the same replication
rate), $g=0.1$ and $m$ as indicated in the figure.}
\label{fig:2}       
\end{figure}

\begin{figure}
\resizebox{0.75\columnwidth}{!}{%
  \includegraphics{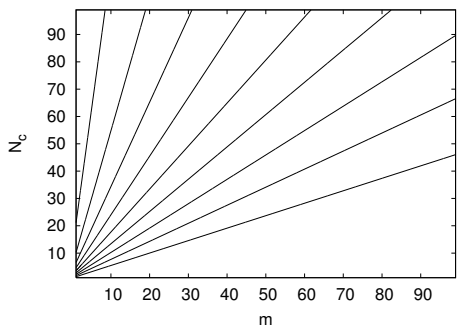}
}
\caption{Critical value of  the vesicle capacity as function of the ratio between 
vesicle  and template generation times  for $\tau=0$  and  (left to right) 
$g=0.1,0.2, \ldots, 0.9$. Below $N_c$, the fixation of
one of the template types via random drift bars coexistence.}
\label{fig:3}       
\end{figure}

Here we will focus on the fraction $\Omega$ of vesicles in the stationary regime 
that carry copies of the two
template types regardless of their redundancies, so that $\Omega = 1 - Y_{N}^{\infty} - Y_0^\infty$.

As pointed out before, in the case of degenerate templates $\tau=0$ the sole obstacle to
coexistence is the effect of random drift, which is intensified for small vesicle capacities.
This result is illustrated in Figure \ref{fig:2}, where we  show also the effect of increasing
the ratio $m$ between the vesicle and template generation times. In this, as well as in
the following figures in which the integer quantities $N$ and $m$ are depicted,
the continuous lines are simply  guides to  the eye. For $N \to \infty$ we find
exceptionally  that $\Omega \to 1$. Two features of this figure are worth emphasizing. First, 
for a fixed value of the cooperation pressure $g$, there is a minimum value of the 
vesicle capacity below which coexistence is unattainable. Second, this threshold value,
 denoted by $N_c$, increases
with increasing $m$, i.e., coexistence is inhibited if the local template populations
evolve faster than the vesicle population. In fact, Figure \ref{fig:3} shows that
$N_c$ increases linearly with $m$ and that the slopes of the straight lines
decrease in a nontrivial way as the cooperation pressure  increases.

\begin{figure}
\resizebox{0.75\columnwidth}{!}{%
  \includegraphics{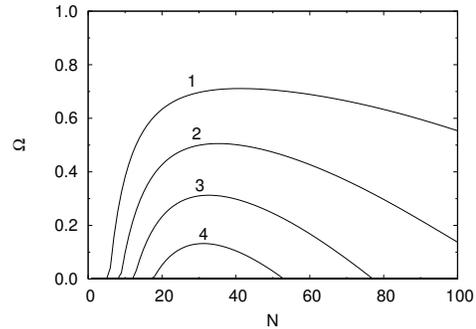}
}
\caption{Fraction of vesicles with the two template types in their compositions 
as function of the vesicle capacity for $\tau=0.1$, $g=0.35$ and $m$ as indicated in the
figure.}
\label{fig:4}       
\end{figure}

\begin{figure}
\resizebox{0.75\columnwidth}{!}{%
  \includegraphics{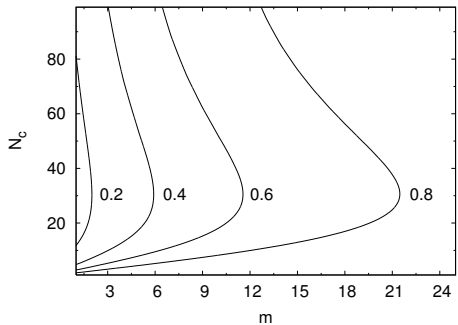}
}
\caption{Critical value of  the vesicle capacity as function of the ratio between 
vesicle and template  generation times  for $\tau=0.1$  and $g$ as
indicated in the figure. For fixed $m$ the two values
of $N_c$ determine the range of vesicle capacity within which coexistence is possible.
}
\label{fig:5}       
\end{figure}

\begin{figure}
\resizebox{0.75\columnwidth}{!}{%
  \includegraphics{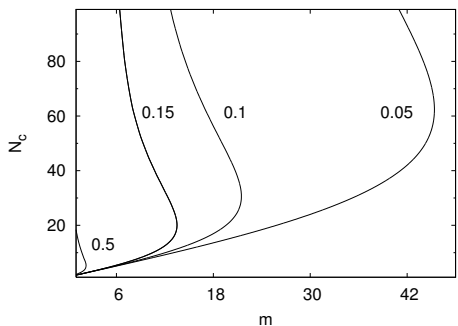}
}
\caption{Critical value of  the vesicle capacity as function of the ratio between 
vesicle and template  generation times  for $g=0.8$  and $\tau$ as
indicated in the figure.}
\label{fig:6}       
\end{figure}

\begin{figure}
\resizebox{0.75\columnwidth}{!}{%
  \includegraphics{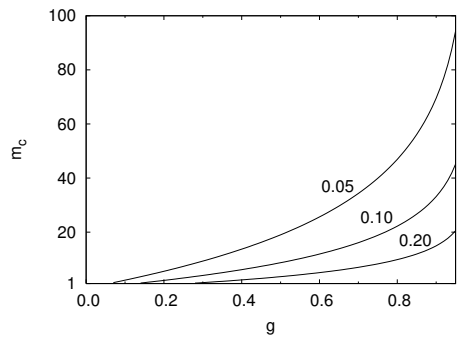}
}
\caption{ Critical value of  the ratio between vesicle and
template  generation times as function of the cooperation pressure 
for $\tau$ as indicated in the figure.  Above  $m_c$ coexistence is impossible.
}
\label{fig:7}       
\end{figure}

We turn now to the analysis of the general case where the template types have different
replication rates, $\tau > 0$. In contrast with the degenerate case discussed before, 
increasing the capacity of the vesicles $N$ will now favor the fixation of the
fittest template, thus inhibiting coexistence.
This is exactly what  Figure \ref{fig:4} depicts: the fraction of vesicles
with the two  template types $\Omega$ is nonzero only for a well-defined range of the
vesicle capacity, which decreases as $m$ increases. We note that because coexistence is
impossible in  the deterministic regime  we have $\Omega = 0$ for $N \to \infty$
regardless of the value of $m$.
Figures \ref{fig:5} and \ref{fig:6} summarize this finding by showing $N_c$  (i.e., the
value of $N$ at which $\Omega$ vanishes) as function of  $m$ for several values 
of the control parameters $\tau$ and $g$. 
As hinted in Figure \ref{fig:4}, there is a
certain value  $m=m_c \left (\tau, g \right )$  
beyond which coexistence is impossible regardless of the vesicle capacity. For
$m < m_c$, we always find two solutions for $N_c$ corresponding to the lower and
higher vesicle capacity compatible with coexistence. The lower bound for $N$ 
changes little with variation of $g$ (see Figure \ref{fig:5}) and is practically
insensitive to variation of $\tau$ (see Figure \ref{fig:6}), provided that the
parameter setting remains within the
coexistence boundary, i.e., $m < m_c$. The upper bound, however, is extremely sensitive
to variation of those parameters.
Finally, Figure \ref{fig:7} illustrates
how $m_c$ depends on the parameters $\tau$ and $g$.

\begin{figure}
\resizebox{0.75\columnwidth}{!}{%
  \includegraphics{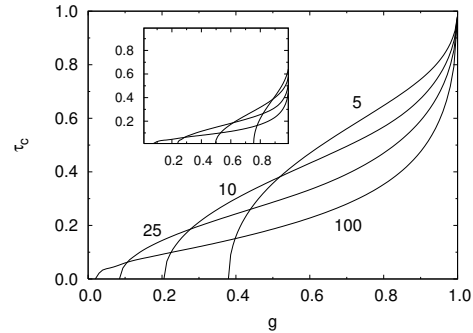}
}
\caption{Critical value of the handicap parameter  as  function of the 
cooperation pressure for $m=1$ (main graph) and $m=5$ (inset). The vesicle capacities
are indicated in the main graph.  
Below  $\tau_c$ coexistence is impossible.
}
\label{fig:8}       
\end{figure}

Up to now we have emphasized the role of the  generation times
of templates and vesicles as  yet another obstacle to the
coexistence of distinct template types, thus generalizing previous approaches
that assumed that template and vesicle populations were updated (this term is appropriate
since in both cases it is assumed that generations do not overlap) with the same
frequency, i.e., $m=1$ \cite{Fontanari}. For completeness, in Figure \ref{fig:8}
we move $m$ to a secondary position and stress the effect of the cooperation pressure
$g$
and handicap $\tau$ on the coexistence of templates $A$ and $B$. These results illustrate
clearly that drift is the main obstacle to coexistence in the case the handicap $\tau$ 
is small, since a large cooperation pressure is needed to retain the two 
template types for small vesicle sizes.  As $\tau$ increases, however, the 
deterministic template competition rapidly takes the lead as the main hindrance to coexistence.
The effect of increasing $m$, as shown in the inset, is to shift non-uniformly
the coexistence lines to higher values of the cooperation pressure.

%
\section{Conclusion} \label{sec:conclusion} 

Compartmentalization of unlinked templates is widely recognized as a necessary step
towards the evolution of cellular life, but it is rarely appreciated that confining
a finite number of templates in a  vesicle actually  creates  a new
obstacle to  coexistence -- the fixation of
one of the template types caused by random drift. Since this disruptive effect is enhanced in
vesicles of low capacity, one should not expect to find such vesicles in a realistic
prebiotic scenario. On the other hand, the risk of vesicles of large capacity
is well-known: the deterministic competition between templates  results in the
fixation of the type with the higher replication rate. So very large vesicles  are not to
be expected in a prebiotic scenario too. 

The range of permitted vesicle capacities depends
on several biologically relevant parameters, and we can speculate on the
values that produce a sensible scenario.
For instance, it is now well established that vesicles spontaneously assembled from fatty
acid micelles \cite{Hanczyc03,Chen04} (see also \cite{Segre01})  grow and  divide
competing for the free micelles in the environment. Hence vesicles can do well
without templates and so the parameter $g$ that appears in the survival probability of
the vesicles, equation (\ref{geom}), should be set to a small value, but not a too
small one since then it would be impossible to compensate for the pressures of drift
and competition.
This observation
immediately excludes low capacity vesicles  from our prebiotic scenario
(see Figure \ref{fig:8}). However, the same figure shows that 
coexistence in high capacity vesicles, say $N=100$,
is possible only if the templates have near degenerate replication rates ($\tau \ll 1$).
Of course, one may object that this is a precarious situation since sooner or later 
a mutant template with a higher replication rate will show up and destroy this 
fragile balance. The answer to this criticism  brings out the main advantage of 
compartmentalization: the mutant (or parasite) appears in a single vesicle which is then 
unlike to pass the extinction stage of the life cycle, since  by definition
the presence of the parasite reduces its chance of  survival, 
equation (\ref{geom}).
The parasite is then quickly eliminated as a result  of the death of the infected vesicle.

Although one might think it is plausible to assume that the templates 
have a much shorter generation time than the vesicles, i.e.,  $m \gg 1$, 
that would prevent coexistence  altogether (see Figure \ref{fig:7}). In addition,
following the arm-chair argument given before, one expects  the number of infected 
vesicles to increase with the number of template replication cycles $m$ 
simply because mutants appear as results of errors during the replication stage. 
So, though counter-intuitive at first sight, we are forced to admit that
the template generation time must be of the same order or even longer 
than that of the vesicles.  Our model, as well as all known package models, is  not 
suitable to describe this  possibility ($m<1$), but it seems clear
from our results that coexistence would be favored in this case. Attempt at modifying
the present analytical formulation to describe the case $m < 1$ is under way.

The ultimate goal of theoretical research on prebiotic evolution is to come up with a 
coherent scenario for the origin of life. Our study supports the view that such a plot
begun with a very large  population of vesicles capable
of template-independent reproduction. The accidental  assimilation of different species of
unlinked templates that happened  to boost the  
reproduction capability of the vesicles  assured then that only vesicles
containing a special kind of templates -- those with near-degenerate
replication rates and long generation times -- would thrive. At this point
the information crisis was overcome and the stage for
a genetic takeover was set: the vesicles would soon loose their ability to reproduce
without the aid of templates, and become thus  mere vehicles or means for template
replication. To refine or reject this scenario is the main theme of theoretical research
on the evolution of life \cite{Szathmary03}.

\bigskip

{\small The work of J.F.F. was supported in part by CNPq and FAPESP, Project No. 04/06156-3.
D.G.M.S. was supported by CNPq.}

\end{document}